\documentclass{ws-ijmpd}

\begin{document}

\markboth{T. Vachaspati}
{Light superconducting strings in the Galaxy}

\catchline{}{}{}{}{}
%
%%%%%%%%%%%%%%%%%%%%%%%%%%%%%%%%%%%%%%%%%%%%%%%%%%%%%%%%%%%%%%%%%%%%

\title{LIGHT SUPERCONDUCTING STRINGS IN THE GALAXY\footnote{Talk given by
T. Vachaspati at the workshop {\it From Quantum to Cosmos: Fundamental
Physics Research in Space}, Washington, D. C., May 2006}}

\author{Francesc Ferrer and Tanmay Vachaspati}

\address{CERCA, Department of Physics, Case Western Reserve University,\\
10900 Euclid Avenue, Cleveland, OH 44106-7079, USA}

\maketitle

%\begin{history}
%\received{Day Month Year}
%\revised{Day Month Year}
%\comby{Managing Editor}
%\end{history}

\begin{abstract}
Observations of the Milky Way by the SPI/INTEGRAL satellite have confirmed the
presence of a strong 511~KeV gamma-ray line emission from the bulge, 
which require an intense source of positrons in the galactic center. These
observations are hard to account for by conventional astrophysical scenarios,
whereas other proposals, such as light DM, 
face stringent constraints from the diffuse gamma-ray
background. Here we suggest that light superconducting strings
could be the source of the observed 511~KeV emission. 
The associated particle physics, at the $\sim$~1~TeV scale, is within
reach of planned accelerator experiments, while the 
distinguishing spatial distribution, 
proportional to the galactic magnetic field, could be mapped by SPI or
by future, more sensitive, satellite missions.

\end{abstract}


\section{Positron Sources in the Galaxy}	
The problem of the birth, propagation and annihilation of 
positrons in the Galaxy has been a major topic of astrophysical investigation,
since the first detection\cite{johnson} of the 511~KeV gamma-ray 
line signature. The SPI instrument on-board ESA's INTEGRAL satellite
has established the presence of a diffuse source of positrons in the 
Galactic Center (GC)\cite{SPI}. The observed photon flux of 
\begin{equation}
9.9^{+4.7}_{-2.1} \times 10^{-4}~{\rm cm}^{-2} {\rm s}^{-1}
\label{f0}
\end{equation} 
with a line-width of about 3~KeV is in good agreement with previous 
measurements\cite{osse}.

For the spatial distribution of the 511~KeV line component, the mapping
results point to an intense bulge emission, better explained by an extended 
distribution than by a point source. Assuming a Gaussian spatial distribution 
for the flux, a full-width at half maximum of $9^\circ$ is indicated. The disk 
component has been either absent, or weakly detected in the initial
results.

The origin of these Galactic positrons remains a mystery. Several
scenarios involving astrophysical sources have been proposed, including 
neutron stars or black holes, massive stars, supernovae, hypernovae, 
gamma ray bursts or cosmic rays\cite{astrof}. However, the fraction of 
positrons produced in such processes is uncertain, and it is unclear that 
the positrons could fill the whole bulge. 

Alternatively, mechanisms associated with the Dark Matter~(DM) at the GC 
have been put forward. If DM is constituted by a light (~MeV) scalar, 
its decay or annihilation could account for the observed 
signal\cite{Boehm:2003bt}. 

The positrons should be injected at non-relativistic energies so that 
the associated bremsstrahlung emission does not 
violate the COMPTEL and EGRET measurements of diffuse radiation from the 
Galactic Center\cite{beacom}. For the DM scenario, this implies that 
the DM particles should be lighter than $\sim$~20-30~MeV. Moreover, inflight 
annihilation of positrons would also overproduce gamma-rays unless the 
positrons are injected at energies below $\sim$~3~MeV\cite{beacom2}, 
thus disfavoring
 some of the scenarios in\cite{astrof} and\cite{Boehm:2003bt}.

We will discuss here the possibility, proposed in\cite{Ferrer:2005xv}, 
that a network of light superconducting strings\cite{Witten:1984eb} 
occurring in particle physics just beyond the standard model 
 could be a source of the galactic positrons. This scenario predicts 
a characteristic positron distribution that could be used to distinguish
this source from the other possibilities. 

Assuming that a tangle of superconducting strings exists 
in the Milky Way, then
the strings are frozen in the  
plasma as long as the radius of curvature is larger than a certain 
critical length scale. If the curvature radius is smaller, the 
string tension wins over the plasma forces and the string
moves with respect to the magnetized plasma.
During the string motion, the loop will cut across the Milky
Way magnetic field, generating current as given by Faraday's
law of induction. 

The current is composed of zero modes of charged 
particles, including positrons, propagating
along the string. The external magnetic field shifts the modes of the 
charge carriers into the bulk and modifies the dispersion relation 
dramatically so that their energy remains below the threshold for 
expulsion\cite{Ferrer:2006ne} (which for the positron zero modes is 511~KeV).
An additional perturbation, like inhomogeneities in the magnetic field,
string motion and curvature, or scattering by counter-propagating 
particles\cite{Barr:1987ij}, ejects the zero modes at the 
threshold of 511~KeV. The ejected positrons will annihilate with the 
ambient electrons, thus emitting 511~KeV gamma rays.

\section{Light superconducting strings in the Galaxy}

The amount of positrons injected in the Milky Way, will
depend on how many strings are injecting positrons per unit volume and
on the output rate of positrons per unit length of string. Let us first 
estimate the density of strings in the Galaxy.

The strings, being superconducting, can sustain currents which couple 
the dynamics of the string network to the Milky Way plasma. The density
of strings will, thus, depend both on the properties of the string
(like its tension, $\mu$, radius of curvature, $R$, and the intensity of
the current being carried, $J$) and of the plasma (like its density,
$\rho$).

The dynamics is determined by comparing the force due to
string tension, $F_s$,  to the plasma drag force, $F_{\rm d}$.
The analysis\cite{Chudnovsky:1986hc} shows that
there is a critical radius of curvature,
\begin{equation}
R_c \sim \frac{\mu}{\sqrt{\rho} J}
\end{equation}
such that the plasma drag can not check the force due to the string tension
when $R < R_c$, and the strings move at relativistic speeds. 
String loops will then emit electromagnetic radiation and eventually dissipate.

On the other hand, less curved strings, i. e. for $R> R_c$, accelerate
under their own tension until they reach a terminal velocity
\begin{equation}
v_{\rm term} \sim \frac{\mu}{\sqrt{\rho} JR}.
\label{vterm}
\end{equation}

In a turbulent plasma, such as in our Milky Way, there is another 
length scale of interest, called $R_*$ ($R_* > R_c$), even when the 
string motion is overdamped. For $R > R_*$, the terminal speed of
the strings is small compared to the turbulence speed of the plasma
and the strings are carried along with the plasma. As the strings
follow the plasma flow, they get more entangled due to turbulent
eddies, and the strings
get more curved until the curvature radius drops below $R_*$. Then
the string velocity is large compared to the plasma velocity, and, 
hence, the strings break away from the turbulent flow. Therefore,
$R_*$ is the smallest scale at which the string network follows
the plasma flow. For $R_* > R > R_c$, the string motion is over-damped
but independent of the turbulent flow. Hence, string curvature on 
these scales is not generated by the turbulence, and we can estimate
the length density of strings in the plasma as $\rho_l \sim {1}/{R_*^2}$.
The scale $R_*$ at which the terminal velocity~(\ref{vterm}) equals the 
turbulent velocity of the plasma, $v_*$, is given by\cite{Chudnovsky:1988cv}:
\begin{equation}
R_* \sim l 
   \left ( \sqrt{\frac{\mu}{\rho}} \frac{1}{e \kappa v_l l} \right )^{4/5}
~ , \ \ 
v_* \sim v_l 
    \left ( \sqrt{\frac{\mu}{\rho}} \frac{1}{e \kappa v_l l} \right )^{1/5},
\label{Rstar}
\end{equation}
where, for convenience, the dimensionless parameter $\kappa$ has been 
introduced via $J \equiv \kappa e \sqrt{\mu}$.

\section{Particle emission by superconducting strings} 

When a string of typical length $R_*$ moves at a velocity $v_*$ with respect
to the Milky Way plasma, it cuts across the 
galactic magnetic field lines and a current is generated on the string
as described by Faraday's law of induction.

\begin{figure}[pb]
\centerline{\psfig{file=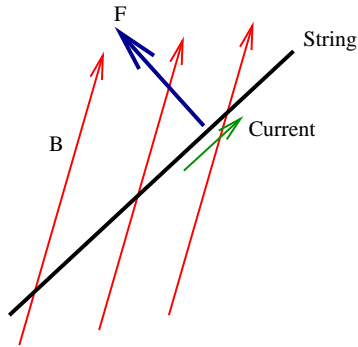,width=4.7cm}}
\vspace*{8pt}
\caption{The charge carriers run along the string, in the presence
of a perpendicular external magnetic field.}
\end{figure}

The same magnetic field that creates the current, changes the dispersion 
relation of the zero modes. In the absence of the external magnetic field,
the zero modes behave as massless particles, with a linear dispersion relation.
The current could increase indefinitely as the zero modes gain momentum.
However, the presence of the external magnetic field changes the dispersion
relation of the zero modes dramatically\cite{Ferrer:2006ne} and it can be
approximated by:
\begin{equation}
\omega_k = m_{e^+} {\rm tanh} \left(\frac{k}{k_*} \right),
\end{equation}
where $k_*$ is a parameter that depends on the magnetic field.
Consequently, the current on the string saturates at $J_{max}=e m_{e^+}$, 
and the positron zero modes have energies approaching from below 
the threshold for expulsion, 511~KeV.

The string will, in general, carry additional zero modes corresponding 
to other heavier particles, say heavy quarks. Then, the total current in 
the string, entering the network dynamics in Eq.~(\ref{Rstar}), will be 
determined by the heavier particles. The positron current, though, will 
still be bounded to be below 511~KeV.

A given charge carrier can, in principle, leave the string once it has enough 
energy. The escape is triggered by several factors such as
string motion and curvature or scattering by counter-propagating 
particles\cite{Barr:1987ij} ($u$ quarks for electroweak 
strings\cite{Ferrer:2005xv}), but in any case, the positrons will be emitted 
at their threshold energy, 511~KeV.

As a piece of string of length $R_*$ cuts through a magnetic field
$B$, it will produce electrons or positrons, with equal likelihood,
at the rate
\begin{equation}
\frac{d N}{dt} \sim e v_* B R_*
\end{equation}
In a volume $V=4\pi L^3/3$, there are of order $L^3/R_*^3$ such pieces 
of string and hence the rate of particle production in the entire volume 
is:
\begin{equation}
\frac{d N_{\rm V}}{dt} \sim e v_* B \frac{L^3}{R_*^2}
\label{dNVdt}
\end{equation}
The current in the positrons will grow at first, but then saturate
at 511 KeV. 
After that, further motion of the string across the galactic magnetic
field will generate positrons that leave the string. So $N_V$ is also the 
number of positrons being produced in the volume $V$ which we denote
by $N_+$. Inserting Eqs.~(\ref{Rstar}) in (\ref{dNVdt})
we get:
\begin{equation}
\frac{d N_+}{dt} \sim 
     10^{42} B_3 \kappa^{7/5} V_1 \mu_1^{-7/10} \rho_{\rm gc}^{7/10} 
        v_{l,6}^{12/5} l_{100}^{-3/5}\;{\rm s}^{-1},
\label{positronrate}
\end{equation}
where we have introduced the parameters describing the
plasma of the spherical region of radius 1 kpc around the galactic center,
$\rho \sim 6 \cdot 10^{-24} \rho_{\rm gc} {\rm gm/cm^3}$,
 $B = B_3 10^{-3}$ G, $v_l=10^6~v_{l,6}$ cm/s 
and $l=100 ~ l_{100}$ pc. The string tension is given by
$\mu = \mu_1 (1~{\rm TeV})^2$.

Although the astrophysical parameters describing the galactic
center are not known very accurately,
assuming equipartition of plasma kinetic energy ($\sim \rho v_l^2$) and 
magnetic energy ($\sim B^2/8\pi$), with $l \sim \rho^{-1/3}$, we find that
$v_{l,6} \sim 100$ and $l_{100} \sim 0.1$ which boosts the estimate in 
Eq.~(\ref{positronrate}) by $10^5$, yielding 
\begin{equation}
\frac{d N_+}{dt} \lesssim 10^{47}~ {\rm s}^{-1}
\label{numericalestimate}
\end{equation} 

This should be compared with the actual 
positron production rate in the galactic center:
\begin{equation}
\frac{d N_+^{obs}}{dt} 
  \sim 1.2 \times 10^{43}\;{\rm s}^{-1}.
\label{observedrate}
\end{equation} 

Comparing Eqs.~(\ref{numericalestimate}) and (\ref{observedrate})
we conclude that light superconducting strings are possible sources
of positrons that lead to the flux of 511~KeV gamma rays observed by 
the INTEGRAL collaboration.

\section{Observational signatures}

We see from Eq.~(\ref{positronrate}) 
that a unique prediction of our scenario is that the gamma ray 
flux is proportional to the strength of the magnetic field in the
Milky Way, with a milder dependence on the plasma density. 
In the disk, $B_3 \sim 10^{-3}$, and we estimate a photon flux 
$\sim 10^{-6}~{\rm cm}^{-2}{\rm s}^{-1}$ in a $16^\circ$ field
of view as in SPI. The target sensitivity of the SPI instrument, once 
sufficient exposure becomes available, is $2\times 10^{-5} ~ 
{\rm cm}^{-2}{\rm s}^{-1}$ at 511~KeV\cite{spi}. This threshold 
is somewhat above what is needed to map the emission from the 
disk in our scenario. 

That the flux should follow the magnetic field is in marked 
contrast with the MeV DM hypothesis. There, the flux follows 
$\rho_{\rm DM}^2$, and a signal from nearby DM dominated regions, 
{\it e.g.} the Sagittarius dSph galaxy, is expected\cite{Hooper:2003sh}.
If superconducting strings source the observed 511~KeV, however,
at most, a flux of $\sim 10^{-7}~{\rm cm}^{-2} {\rm s}^{-1}$ 
in the direction of Sagittarius is expected\cite{Ferrer:2005xv},
some three orders of magnitude fainter than the MeV DM model prediction

The strings are expected to carry additional zero modes apart from the
ones corresponding to $e^\pm$. We expect, thus, the presence of other currents,
each saturated at the mass $m_X$ of the particle in vacuum, which would
potentially result in the ejection of these particles also at threshold,
although the presence of conserved charges might inhibit or delay some of
these processes. For instance, since pion emission cannot 
deplete the baryonic current on the string, 
only at $\sim 1$~GeV energies can 
antiprotons be emitted, leading to another possible signature of 
galactic superconducting strings\cite{Starkman:1996hi}.

Our scenario could also explain the excess of high-energy
positrons in cosmic rays at energies around 10~GeV detected by
the HEAT balloon experiment\cite{HEAT}. Since the positron current cannot
build up beyond 511~KeV in the presence of the external magnetic field,
additional heavy charged fermions would be responsible, after decaying
or annihilating with ambient particles in the plasma, for 
the positrons in the 10~GeV energy range\cite{Ferrer:2005xv}.

\section{What can NASA do to check these predictions}

The experimental results obtained by the SPI/INTEGRAL collaboration, have
confirmed the puzzle of the positron injection in the Galactic bulge and
sparked, in the few years since the publication of the first results, 
a plethora of possible explanations.

The distinguishing feature of our scenario is the spatial distribution tracking
the magnetic field intensity. SPI will not be able to attain the sensitivity
required to observe the emission from the galactic disk in our model. An
improvement of roughly an order of magnitude in the sensitivity would suffice
to pursue this task\footnote{Other sources, e.g. cosmic 
rays, could contribute to the emission from the disk at a 
level that could be mapped
by SPI}. This order of magnitude improvement would also test predictions from 
other scenarios. For instance, the signal from nearby DM clumps expected in
light DM scenarios could be unveiled or, else, the models would be disproved
(barring astrophysical uncertainties in the region of the clumps).

It is noteworthy, that the analysis of complementary data coming from older
satellites like COMPTEL and EGRET, provides some of 
the most stringent constraints for
all the scenarios\cite{beacom,beacom2}. In this respect, 
light superconducting strings remain a 
viable proposal, since the positrons are emitted at threshold, well below
the $\sim$~3~MeV bounds from inflight annihilation. In should be stressed that
these bounds require the knowledge of the diffuse gamma-ray flux in the
galactic center
to a great accuracy. With the data at hand, extrapolations of data at 
different energies and from different regions 
are necessary, which add uncertainty to the bounds. 
The forthcoming GLAST satellite, partially funded by NASA, 
is better suited to energies above 1~GeV. The task remains
to get more precise data at lower energies. SPI is already contributing to this
effort, but a more precise experiment would make a difference.

\section*{Acknowledgments}

%\begin{thebibliography}{000} %for 3 digits

\end{document}